\documentclass[12pt]{iopart}
\begin{document}
\title{Lagrangian of the quasi-rigid extended charge.}
\author{Rodrigo Medina}
\address{Instituto Venezolano de Investigaciones Cient\'{\i}ficas, 
Apartado 21827, Caracas 1020A, Venezuela}
\ead{rmedina@ivic.ve}

%%%%% Abstract %%%%%%
\begin{abstract}
 It is proposed a Lagrangian for the quasi-rigid extended charged particle,
which consists of a bare point particle term plus the standard
electromagnetic minimal coupling. The quasi-rigid motion is imposed as
a constraint.  The extension of the particle and the quasi-rigid motion appear
 inside the current density. The Lorentz contraction of
the extended particle makes the interaction term dependent on the acceleration.
This dependence produces the additional terms in the equations of motion that
are necessary for the proper energy and momentum conservation, and that were
previously identified as the inertial effects of stress. The momentum of stress
is obtained as an explicit function of the electromagnetic field.
\end{abstract}
\submitto{\JPA}
\maketitle

%%%%% Section I
\section{Introduction}

In a previous article \cite{Medina1} the electrodynamics of a classical
extended charge was studied from various points of view. It was assumed that
the particle follows a quasi-rigid motion. The main results of that paper
are the following.
\begin{itemize}
\item[(1)] The $4/3$ problem is solved.
As it is well known that the momentum and energy of the electromagnetic fields
that surround a spherically symmetric charge distribution moving with velocity
$\bi{v}$ are respectively
$\frac{4}{3}U_{\rm e}\gamma\bi{v}/c^2$ and
 $U_{\rm e}\gamma [1+\frac{1}{3}(v/c)^2]$, were $U_{\rm e}$ is the
electrostatic energy and $\gamma=[1-(v/c)^2]^{-1/2}$. These values of
energy and momentum do not form
a 4-vector and seem to contradict the mass-energy equivalence. One should
expect that the mass of the dressed particle be the bare mass $m_0$
plus the electromagnetic mass $U_{\rm e}/c^2$.
It was shown that everything fits in place,
once one considers the inertial effects of the stress that develops inside
the particle to balance the electrostatic repulsion.
 For a particle moving with no acceleration
those effects can be included in a negative pressure contribution
to the mass $m_P$,
\begin{equation}\label{PressureMass}
 m_P=-\frac{1}{3c^2}U_{\rm e}\,.
\end{equation}

So the mass of the dressed particle (bare + stress + bounded fields) has the
expected value $m_0+U_{\rm e}/c^2$. The relevance of the inertial effects of
stress is also discussed in \cite{Medina3}. In this respect it worth
mentioning the Boyer \cite{Boyer} and Rohrlich \cite{Rohrlich1} controversy
that was not discussed in the previous paper \cite{Medina1}. Boyer claimed
that it was wrong to modify the electromagnetic energy-momentum tensor
$T^{\mu\nu}$ in order to eliminate the extra $1/3$ term. Rohrlich maintained
the opposite opinion. The controversy is settled in favour of Boyer. Rohrlich
assumes, in his reply to Boyer, that the interaction that equilibrates the
electrostatic repulsion enters in the equation as a force density, while
actually it is the stress of the particle.  The fact that the energy and the
momentum of the fields do not form a 4-vector is due to the fact that not
only the particle is stressed, but also the fields that surround it. $T^{\mu
\nu}$ should be named more properly as the energy-momentum-stress tensor.
\item[(2)] It is  established the role played by the various
components of the  $T^{\mu\nu}$ in the energy and momentum conservation.
The fields produced by the particle have two components. The radiated field,
that decays as $1/r$ and the bound field that decays as $1/r^2$. As the
tensor $T^{\mu\nu}$ is quadratic in the field there are three components
$T_{BB}$, $T_{BR}$ and $T_{RR}$. The radiation term $T_{RR}$  gives the
energy and momentum of the radiated fields. The term $T_{BB}$ corresponds
to the field bounded to the particle and gives the electromagnetic
contributions to the dressed particle. The cross term $T_{BR}$ is also bounded
to the particle but it only exists as long as there is radiation. The radiation
reaction contains a term that corresponds to radiated momentum, but also
a term that corresponds to the cross term. This cross term behaves as reservoir
of energy and momentum.

 This splitting of $T^{\mu\nu}$ was found long ago by Teitelboim
\cite{Teitelboim1}. He obtained the Lorentz-Abraham-Dirac (LAD) radiation
reaction formula for a point charge using the retarded fields. In the cited
paper Teitelboim used the energy-momentum conservation law of the field. 
The energy-momentum contribution to the dressed particle was calculated in
the reference frame of instantaneous rest of the particle, and then transformed
to the laboratory frame assuming that energy and momentum form
a 4-vector. In this way the stress contribution is suppressed. As
everybody else he disregarded the stress contribution to the mass of the
particle. As a result the expected mass of the dressed particle was
obtained $(m_0 + e^2/(2\epsilon c^2))$. In a later paper \cite{Teitelboim2}
determined the radiation reaction by calculating directly the force produced
by the self-fields. He obtained the LAD formula, but now, of course, the
electromagnetic mass included the stress contribution and was
 $2e^2/(3\epsilon c^2)$. About this discrepancy he wrote in a note:
``However, there is actually no difference between the two expressions, since
the limit $\epsilon\to 0$ is to be taken.'' One cannot agree with
that. Actually, the difference  shows that the stress of the field makes a
real contribution to the momentum of the particle.

\item[(3)] It is found an exact formula for the radiation reaction of the
extended particle. The self-force is given as an integral over the retarded
accelerations. In the $R \to 0$ limit the LAD result is recovered.

\item[(4)] It is found that the solutions of the integro-differential
equation of motion that results when the exact radiation reaction formula
is used do not violate causality or run away, provided that the mass of
matter $m_0 + m_P$ is positive. When the condition
\begin{equation}\label{MassCondition}
m_0 > \frac{U_e}{3c^2}
\end{equation}
is not verified the causality is violated and run-away solutions appear.
Therefore the point particle is inconsistent in classical electrodynamics,
as $\lim_{R\to 0}U_e =+ \infty$. That is, the limit $R\to 0$ is not physical
as it is interior to a non-physical region. The classical mass
renormalization is also inconsistent. It is impossible to verify
(\ref{MassCondition}) and to go to $-\infty$ at the same time.

\item[(5)] It is shown that for a physical point particle that verifies
 (\ref{MassCondition}) the exact radiation reaction formula reduces to the
Rohrlich formula \cite{Rohrlich2}. A physical point particle is a particle
whose radius is much smaller that any other distance
in the problem, in particular than the wave-length of the fields that it
itself generates. The Rohrlich formula is like the LAD formula but replacing
the acceleration $\bi{a}$ by the external force $\bi{F}$ divided by the
dressed mass $m$.

\item[(6)] Finally it is shown that the radiated power of the physical point
charge is not given by the Larmor formula, which is valid for $R\to 0$, but
by a modified one.
\end{itemize}
\noindent
The conclusion of the previous paper \cite{Medina1} is that  the dynamics of
a classical quasi-rigid extended charge, including self-interactions, is,
unlike that of a point charge, perfectly
consistent and conforming with causality and conservation of energy
 and momentum.

The extended particle should have some kind of structure that generates
the stress that balances the electrostatic repulsion.  Nevertheless if a
particle of radius $R$ moves with an acceleration which is small in comparison
with $c^2/R$, it continuously keeps its spherical shape as seen in the
reference frame of instantaneous rest. In this quasi-rigid motion the
internal dynamics is frozen, so the particle moves as its only three
degrees of freedom were the coordinates of its centre. The quasi-rigid
motion corresponds to a constraint that eliminates the internal degrees
of freedom. It has been shown \cite{Medina2} that the pressure mass is the
same for any elasticity model with spherical symmetry, so the elastic
properties should not be relevant in the quasi-rigid limit.

Here we show that the mechanics of the quasi-rigid extended particle
can be obtained from the standard electromagnetic Lagrangian with the
minimal coupling. The peculiarity of the extended particle is that the
current density depends on acceleration. This is due to the fact that
as the speed changes the Lorentz contraction changes and that, therefore,
different points of the particle should move with different velocities. 
Such dependence on the acceleration is unavoidable. Higher order Lagrangians
are rare. One may imagine elasticity models of the particle that correspond
to first order Lagrangians. The interaction which is proportional to the
velocity of each point will also be of first order. It is the quasi-rigid
constraint, that makes the motion of each point of the particle a function
of the motion of its centre, what produces the acceleration dependent
interaction. That is the price one has to pay for having eliminated the internal
degrees of freedom.

%%%%%% Section II 
\section{The Lagrangian}
We will assume that only electromagnetic forces are acting on the
particle, but that, in addition to the fields generated by the particle
itself, there are also those due to some external current density
$j^{\mu}_{\rm ex}$. We will use Gauss electromagnetic units and the metric tensor $g^{\mu\nu}$
 with positive trace.  We will call $z^{\mu}$ a generic point of four-space
and $x^{\mu}$ the coordinates of the centre of the particle, both in the
laboratory frame. The origin of the instantaneous rest frame  will
be the centre of the particle while $y^{\mu}$ will be the generic point
and $y=|\bi{y}|$.
We will assume that the particle has a non-zero charge $q$ and that
in the instantaneous rest frame it has a constant relative charge density
$g(y)$, which has spherical symmetry and is normalized to 1

\begin{equation}\label{Normalization}
\int\! d^3y\, g(y)=1 .
\end{equation}

Each point of the particle can be labelled with its position in the
instantaneous rest frame, $y^{\mu}$. The quasi-rigid motion is defined
in reference \cite{Medina1} so that the position
of any point of the particle is given at any time  by

\begin{equation}\label{QuasiRigid}
\bi{x}(\bi{y},t)= \bi{x}(t) + \bi{y} + \delta\bi{y} ,
\end{equation}
where $\delta\bi{y}$ is the Lorentz contraction
\begin{equation}\label{Contraction}
\delta\bi{y} = (\gamma^{-1} -1)(\bi{y}\cdot\hat{\bi{v}})\hat{\bi{v}} .
\end{equation}
The quantity $\gamma$ is calculated with the velocity of the centre $\bi{v}$ and
$\hat{\bi{v}}$ is the unit vector in the direction of $\bi{v}$. The actual
motion of a point of the particle differs from the expression (\ref{QuasiRigid})
by a term of order $yaR/c^2$, where $a$ is the acceleration.

The expression (\ref{QuasiRigid}) can be inverted,
\begin{equation}\label{QuasiRigidInv}
\bi{y}= \bi{z}-\bi{x}+(\gamma -1)(\bi{z}-\bi{x})\cdot\hat{\bi{v}}
\,\hat{\bi{v}}.
\end{equation}

Because $\delta\bi{y}$ depends on $\bi{v}$, different points of the particle
have different velocities as the particle is accelerated.

\begin{eqnarray}
\bi{v}(\bi{y},t)&=\frac{\partial\bi{x}(\bi{y},t)}{\partial t}
\\
&=\bi{v}(t)+ \delta\bi{v}(\bi{y},t) ,
\end{eqnarray}
where
\begin{eqnarray}
\delta v_i&=\frac{\partial\delta y_i}{\partial t}
\\
\label{DeltaV}
&=a_j\frac{\partial\delta y_i}{\partial v_j} .
\end{eqnarray}

In the instantaneous rest frame the charge density is $\rho = qg(y)$
and the current density  vanishes. In the laboratory frame the charge density
is
\begin{eqnarray}
\label{Rho}
\rho(z^\mu) &= q\gamma g(|\bi{z}-\bi{x}+ (\gamma -1)(\bi{z}-\bi{x})\cdot
\hat{\bi{v}}\hat{\bi{v}}|)
\\
&=q\int\! d^3y\, g(y)\delta(\bi{z}-\bi{x}-\bi{y}-\delta\bi{y})
\end{eqnarray}
and the current density is
\begin{eqnarray}
\label{Jota}
\bi{j}(z^\mu) &=\rho(z^\mu) (\bi{v}+\delta\bi{v})
\\
&=q\int\! d^3y\, g(y)\delta(\bi{z}-\bi{x}-\bi{y}-\delta\bi{y})
(\bi{v}+\delta\bi{v}) .
\end{eqnarray}
In both expressions we have used the fact that $d^3y=\gamma d^3z$.
The definitions (\ref{Rho}) and (\ref{Jota}) are consistent with the
charge conservation $\partial_{\mu}j^{\mu}=0$.

We can now write down the Lagrangian we propose, namely
\begin{equation}\label{Lagrangian}
L(\bi{x},\bi{v},\bi{a},t,A^\mu)=
-\frac{1}{16\pi}\int\!d^3z\, F^{\mu\nu}F_{\mu\nu} - m_0c^2\gamma^{-1}
+\frac{1}{c}\int\! d^3z(j^\mu+j^\mu_{\rm ex})A_\mu .
\end{equation}

The first term is the Lagrangian of electromagnetic fields, the second
is the Lagrangian of a bare point particle and the third is the
standard electromagnetic coupling. The fact that the particle has extension
and that its motion is quasi-rigid appears in the current density $j^\mu$. The
dependence on the acceleration $\bi{a}$ is in $\delta\bi{v}$. Using the
expressions for the current density, the interaction term of the Lagrangian
can be written as,

\begin{equation}\label{Interaction}
\fl
L_{\rm I}=\frac{q}{c}\int\! d^3y\, g(y)(\bi{v}+\delta\bi{v})
\cdot\bi{A}(\bi{x}+\bi{y}+\delta\bi{y},t)
-q\int\! d^3y\, g(y)\phi(\bi{x}+\bi{y}+\delta\bi{y},t) .
\end{equation}

 It is obvious that this Lagrangian yields the correct Maxwell's equations. 
We will show that it also gives the correct equations of motion of the
particle, but before we will in the next section recall how to handle
Lagrangians that depend on acceleration.

%%%%%% Section III 
\section{Acceleration dependent Lagrangian}
The treatment of higher order Lagrangians was developed by Ostrogradsky
in the middle of XIX century \cite{Ostrogradsky}. The Hamiltonian approach
was made by Govaerts and Rashid \cite{Govaerts}. We resume here the results
for an acceleration dependent Lagrangian $L(q,\dot{q},\ddot{q},t)$. The
conjugate momentum of $q_i$ is
\begin{equation}\label{ConjugateMomentum}
p_i= \frac{\partial L}{\partial \dot{q}_i}-\frac{d}{dt}\frac{\partial L}
{\partial \ddot{q}_i} .
\end{equation}

The Euler-Lagrange equations of motion are
\begin{equation}\label{EL}
\frac{d p_i}{dt}=\frac{\partial L}{\partial q_i} .
\end{equation}

The quantity $\frac{\partial L}{\partial \ddot{q}_i}$ behaves as the
conjugate momentum of $\dot{q}_i$, so the Hamiltonian is

\begin{equation}\label{Hamiltonian}
H=\sum_i \dot{q}_i p_i +
 \sum_i \ddot{q}_i\frac{\partial L}{\partial \ddot{q}_i}
-L .
\end{equation}

Finally the evolution of $H$ is given by
\begin{equation}\label{HEq}
\frac{d H}{dt}=-\frac{\partial L}{\partial t} .
\end{equation}

%%%%%% Section IV 
\section{Equations of motion}
In this section we obtain the equations of motion of the Lagrangian
(\ref{Lagrangian}).  The conjugate momentum of $\bi{x}$ is
\begin{equation}\label{ConjuagateMomentum2}
p_i=\frac{\partial L}{\partial v_i}-\frac{d}{dt}
\frac{\partial L}{\partial a_i} .
\end{equation} 

From (\ref{Interaction}) and (\ref{DeltaV}) we obtain
\begin{eqnarray}
\frac{\partial L}{\partial a_i} &= \frac{q}{c}\int\! d^3y\, g(y)
\frac{\partial\delta v_j}{\partial a_i}A_j  \\
  &= \frac{q}{c}\int\! d^3y\, g(y)\frac{\partial\delta y_j}{\partial v_i}A_j ,
\end{eqnarray}
and
\begin{equation}
\fl
\frac{d}{dt}\frac{\partial L}{\partial a_i} =
\frac{q}{c}\int\! d^3y\, g(y)\frac{\partial\delta y_j}{\partial v_i}
\Big[\frac{\partial A_j}{\partial t}+(v_k+\delta v_k)\frac{\partial A_j}
{\partial x_k}\Big]+ \frac{q}{c}\int\! d^3y\, g(y)a_k
\frac{\partial^2\delta y_j}{\partial v_k\partial v_i}A_j .
\end{equation}

\noindent
On the other hand from (\ref{Lagrangian}), (\ref{Interaction}) and
 (\ref{DeltaV}) we get
\begin{equation}
\eqalign
{
\fl
\frac{\partial L}{\partial v_i}= m_0\gamma v_i + \frac{q}{c}\int\! d^3y\, g(y)
A_i + \frac{q}{c}\int\! d^3y\, g(y)\frac{\partial\delta v_j}{\partial v_i}A_j
\cr
+\frac{q}{c}\int\! d^3y\, g(y)\frac{\partial\delta y_j}{\partial v_i}\Big[
(v_k+\delta v_k)\frac{\partial A_k}{\partial x_j}-c\frac{\partial\phi}
{\partial x_j}\Big] .
}
\end{equation}
\noindent
The conjugate momentum is then
\begin{equation}\label{ConjugateMomentum3}
p_i=m_0\gamma v_i + \frac{q}{c}\int\! d^3y\, g(y)A_i +
q\int\! d^3y\, g(y)\frac{\partial\delta y_j}{\partial v_i}\Big[
\bi{E}+c^{-1}(\bi{v}+\delta\bi{v})\times\bi{B}\Big]_j ,
\end{equation}
where $\bi{E}$ and $\bi{B}$ are the electrical and magnetic fields respectively.

The first term of (\ref{ConjugateMomentum3}) is the momentum of the
bare particle, the second term is the usual vector potential contribution
which is also present for the point particle, but in this case it is averaged
over the whole particle. The last term is
distinctive of the extended quasi-rigid particle. So, we will define
the momentum of constraint as
\begin{equation}\label{ConstraintMomentum}
p_{Ci}=q\int\! d^3y\, g(y)\frac{\partial\delta y_j}{\partial v_i}\Big[
\bi{E}+c^{-1}(\bi{v}+\delta\bi{v})\times\bi{B}\Big]_j .
\end{equation}

 The largest contribution to the constraint momentum comes from the
electrostatic repulsion of the charges of the particle. We will call
momentum of matter $\bi{p}_M$ the sum of the bare momentum plus the
constraint momentum
\begin{equation}
\bi{p}_M=m_0\gamma\bi{v}+\bi{p}_C .
\end{equation}
With these definitions the conjugate momentum is
\begin{equation}
\bi{p}=\bi{p}_M + \frac{q}{c}\int\! d^3y\, g(y)\bi{A} .
\end{equation}

The equation of motion of $\bi{p}_M$ is obtained from (\ref{EL})
\begin{eqnarray}
\dot{\bi{p}}_M&=\dot{\bi{p}}-\frac{q}{c}\frac{d}{dt}\int\!d^3y\, g(y)\bi{A}\\
&=\frac{\partial L}{\partial \bi{x}} -
\frac{q}{c}\int\!d^3y\, g(y)\frac{d}{dt}\bi{A}\\
&=\frac{q}{c}\int\!d^3y\, g(y)\Big[\nabla(\bi{v}+\delta\bi{v})\cdot\bi{A}-c
\nabla\phi-\frac{\partial\bi{A}}{\partial t}-(\bi{v}+\delta\bi{v})\cdot
\nabla\bi{A}\Big]\\
&=q\int\!d^3y\, g(y)\Big[\bi{E}+c^{-1}(\bi{v}+\delta\bi{v})\times\bi{B}
\Big]\\
\label{EquationOfMotion}
&=\int\!d^3z\, (\rho\bi{E}+c^{-1}\bi{j}\times\bi{B}) .
\end{eqnarray}

To call $\bi{p}_M$ the momentum of matter is justified by the fact
that its time derivative is the integral of the force density. Note
that it is different from the dressed particle momentum, which in addition
includes the momentum of the fields that surround the particle. The equation
of motion
(\ref{EquationOfMotion}) is similar to the equation (14) of reference
\cite{Medina1}, but there instead of $\bi{p}_C$ one has the momentum of stress
$m_P\gamma\bi{v}$. In references \cite{Medina1} and \cite{Medina2} the mass
of stress $m_P$ is calculated from the stress tensor of the particle, and
the stress is determined from the stability condition of the particle. Instead
the expression (\ref{ConstraintMomentum}) gives the momentum of stress as an
explicit function of the fields.

%%%%%% Section V 
\section{Energy equation}
The energy is  obtained from (\ref{Hamiltonian}),
\begin{eqnarray}
E &= \bi{v}\cdot\bi{p} + \bi{a}\cdot\frac{\partial L}{\partial\bi{a}} -L\\
\label{Energy}
&=m_0c^2\gamma + \bi{v}\cdot\bi{p}_C + q\int\! d^3y\, g(y)\phi .
\end{eqnarray}

The first term is the bare particle energy, the last one is the electrostatic
potential energy and the second term is a contribution due to the constraint.
As for the momentum we define the energy of matter as the sum of the bare
particle contribution plus the constraint contribution
\begin{equation}\label{MatterEnergy}
E_M=m_0c^2\gamma +\bi{v}\cdot\bi{p}_C .
\end{equation}
Note that $E_M$ and $\bi{p}_M$ do not form a 4-vector.

The time evolution of energy is obtained using (\ref{HEq})
\begin{eqnarray}
\frac{dE_M}{dt} &= -\frac{\partial L}{\partial t}-q\int\!d^3y\, g(y)
\frac{\partial\phi}{\partial t}\\
&=q\int\!d^3y\, g(y)\Big[ -c^{-1}(\bi{v}+\delta\bi{v})\cdot\frac{\partial\bi{A}}
{\partial t}+\frac{\partial\phi}{\partial t}-\frac{d\phi}{d t}\Big]\\
&=-q\int\!d^3y\, g(y)(\bi{v}+\delta\bi{v})\cdot\big(c^{-1}\frac{\partial\bi{A}}
{\partial t}+ \nabla\phi\big)\\
&=q\int\!d^3y\, g(y)(\bi{v}+\delta\bi{v})\cdot\bi{E}\\
\label{EEquation}
&=\int\!d^3z\, \bi{j}\cdot\bi{E} .
\end{eqnarray}

The time derivative of $E_M$ is the integral of power density. This last
equation corresponds to the equation (17) of \cite{Medina1}. The pressure
contribution to the energy that appears there can be written as
$m_P\gamma v^2$, which is consistent with the constraint term of (\ref{Energy}).

%%%%%% Section VI
\section{Identity of the stress momentum  and the constraint momentum}
In this section we will calculate the constraint momentum with the same
approximations that were used in \cite{Medina1}, that is: 1) the size of the
particle is small in comparison with the external currents, so the external
fields can be considered constant inside the particle; 2) the dependence on
acceleration will be neglected. With these conditions the only contributions
to the constraint momentum in (\ref{ConstraintMomentum}) come from the
electrostatic  self-field. In the rest frame the magnetic self-field
vanishes, while the electric field is
\begin{equation}\label{RestE}
\bi{E}=q\frac{Q(y)}{y^2}\hat{\bi{y}}
\end{equation}
where $Q(y)$ is
\begin{equation}
Q(y)=\int_0^y dy\, 4\pi y^2 g(y) .
\end{equation}

In the laboratory frame, expressed in terms of the coordinate of the
rest frame $\bi{y}$, the electric and magnetic self-fields are
\begin{equation}\label{LabE}
\bi{E}=q\frac{Q(y)}{y^2}[\gamma\hat{\bi{y}} + (1-\gamma)
(\hat{\bi{y}}\cdot\hat{\bi{v}})\hat{\bi{v}}]
\end{equation}
and
\begin{equation}\label{LabB}
\bi{B}=\frac{\gamma q}{cy^2}Q(y)\bi{v}\times\hat{\bi{y}} \,.
\end{equation}

As $\bi{a}=0$ then $\delta\bi{v}=0$, so the bracket in
(\ref{ConstraintMomentum}) is
\begin{equation}\label{braket}
\bi{E}+c^{-1}\bi{v}\times\bi{B}=q\frac{Q(y)}{y^2}[\gamma^{-1}\hat{\bi{y}}+
(1-\gamma^{-1})(\hat{\bi{y}}\cdot\hat{\bi{v}})\hat{\bi{v}}] \,.
\end{equation}
On the other hand
\begin{equation}\label{DdyDV}
\frac{\partial\delta y_j}{\partial v_i}=-\frac{yv}{c^2(\gamma^{-1}+1)}
[(\gamma-1)(\hat{\bi{y}}\cdot\hat{\bi{v}})\hat{v}_i\hat{v}_j+
\hat{y}_i\hat{v}_j+(\hat{\bi{y}}\cdot\hat{\bi{v}})\delta_{ij}] \,.
\end{equation}
Therefore, from (\ref{braket}) and (\ref{DdyDV})
\begin{equation}\label{integrand}
\frac{\partial\delta y_j}{\partial v_i}[\bi{E}+c^{-1}\bi{v}\times\bi{B}]_j=
-\frac{qQ(y)}{c^2y}[(\gamma-1)(\hat{\bi{y}}\cdot\hat{\bi{v}})^2v_i+
(\bi{v}\cdot\hat{\bi{y}})\hat{y}_i] .
\end{equation}

As the particle has spherical symmetry, in doing the integral of
(\ref{ConstraintMomentum}) one can first integrate the solid angle, and then
the radial coordinate. The spherical average of (\ref{integrand}) is
readily obtained using the fact that
\begin{equation}
\frac{1}{4\pi}\int\! d\Omega\, \hat{y}_i\hat{y}_j= \frac{1}{3}\delta_{ij} \,.
\end{equation}
The average is
\begin{equation}\label{average}
\frac{1}{4\pi}\int\! d\Omega\, \frac{\partial\delta y_j}{\partial v_i}
[\bi{E}+c^{-1}\bi{v}\times\bi{B}]_j= -\frac{qQ(y)\gamma}{3c^2y}v_i
\end{equation}
and therefore the constraint momentum is
\begin{equation}\label{ConstraintMomemntum2}
\bi{p}_C=-\frac{q^2\gamma}{3c^2}\int\!d^3y\,\frac{g(y)Q(y)}{y}\,\bi{v} .
\end{equation}

The integral in (\ref{ConstraintMomemntum2}) is proportional to the
 electrostatic energy
\begin{eqnarray}
\int\!d^3y\,\frac{g(y)Q(y)}{y} &= \int\!dQ\,\frac{Q(y)}{y}\\
&=\frac{Q(y)^2}{2y}{\bigg\vert}^{\infty}_{0} + \frac{1}{2}\int\!dy\,\frac{Q(y)^2}{y^2}\\
&= \frac{1}{8\pi q^2}\int\!d^3y\,E^2\\
&= q^{-2}U_e .
\end{eqnarray}

The constraint momentum is then
\begin{equation}
\bi{p}_C=m_P\gamma\bi{v}
\end{equation}
where $m_P$ is given in (\ref{PressureMass}). It is exactly the same
expression of the momentum of stress that was given in 
\cite{Medina1,Medina2}. The energy of matter $E_M$ is also
the same that appears in equation (17) of \cite{Medina1},
\begin{eqnarray}
E_M &= m_0c^2\gamma + m_P\gamma v^2 \\
&= (m_0 + m_P)c^2\gamma - m_Pc^2\gamma^{-1} .
\end{eqnarray}

%%%%%% Section VII 

\section{Conclusion}
We have shown that the standard electromagnetic Lagrangian with
minimal coupling $j^\mu A_\mu$ yields the proper behaviour of the quasi-rigid
extended particle. The internal degrees of freedom are not included; instead
the quasi-rigid motion is imposed as a constraint. The velocity of different
parts of the particle are different when the particle is accelerated, so
the current density $\bi{j}$ and the Lagrangian depend on acceleration.
This fact produces additional terms in the momentum and energy of the
particle that are the same that were found in the previous work
\cite{Medina1} to be the inertial effects of stress. These additional terms
exactly cancel the additional terms in the energy and momentum of the
self-fields that surround the particle, and therefore the  dressed particle
(bare particle + constraint + surrounding fields) has a standard
momentum-energy 4-vector corresponding to the expected mass $m_0+U_e/c^2$.
All the results of \cite{Medina1}, in particular the correct
radiation reaction formula, are consistent with the present Lagrangian
formulation.

To have found a proper Lagrangian theory is the first step
towards the quantization of the extended quasi-rigid particle, but the
quantization of acceleration-dependent Lagarangians is not straightforward.
A possible path that may be followed in order to achieve this goal could be
to convert the Lagrangian to a first order one by considering the velocity
$\bi{v}$ as a generalized coordinate independent from $\bi{x}$ and to impose
the condition $\dot{\bi{x}}=\bi{v}$ as a constraint by means of Lagrange's
multipliers. Such singular Lagrangian could be quantized using Dirac's
method \cite{Govaerts}.

%%%% Acknowledgment
\ack
I wish thank Dr. Victor Villalba and Dr. Jorge Stephany for many enlightening
discussions.

\vfill
%%%%%% References

% Referencias para  Journal of Physics A

\end{document}